\documentclass[aps,prb,twocolumn]{revtex4}
\usepackage{epsfig}
\usepackage{graphicx}
\begin{document} 
\title{Self-energy corrections in an antiferromagnet --- \\
interplay of classical and quantum effects on quasiparticle dispersion}
%in the $t-t'-U$ model}
\author{Pooja Srivastava and Avinash Singh}
\email{avinas@iitk.ac.in} 
\affiliation{Department of Physics, Indian Institute of Technology Kanpur - 208016}
\begin{abstract}
Self-energy corrections due to fermion-magnon interaction 
are studied in the antiferromagnetic state of the $t-t'-t''$ Hubbard model
within the noncrossing approximation in the full $U$ range
from weak to strong coupling.
The role of classical (mean-field) features of fermion and magnon dispersion,
associated with finite $U,t',t''$,
are examined on quantum corrections to quasiparticle energy, weight,
one-particle density of states etc.
A finite-$U$ induced classical dispersion term, absent in the $t-J$ model,
is found to play an important role in suppressing
the quasiparticle weight for states near ${\bf k}=(0,0)$, as seen in cuprates.
For intermediate $U$, the renormalized AF band gap is found to be
nearly half of the classical value,
and the weak coupling limit is quite non-trivial due to
strongly suppressed magnon amplitude. For finite $t'$,
the renormalized AF band gap is shown to vanish at a critical
interaction strength $U_c$, yielding a spin fluctuation driven
first-order AF insulator - PM metal transition.
Quasiparticle dispersion evaluated with the same set
of Hubbard model cuprate parameters, 
as obtained from a recent magnon spectrum fit,
provides excellent agreement with
ARPES data for $\rm Sr_2 Cu O_2 Cl_2 $.
\end{abstract}
\maketitle

\section{Introduction}
Driven by the intensive effort to understand the fascinating properties
of doped cuprates, substantial theoretical progress has been made
in recent years in the area of strongly correlated electron systems.
In particular, the study of quantum spin fluctuations in the
antiferromagnetic (AF) state
and the renormalization of quasiparticle properties of doped carriers
due to coupling with spin fluctuations has attracted much attention
in the context of antiferromagnetic spin correlations,
anomalous normal state properties, spin-fluctuation mediated pairing,
optical conductivity etc.\cite{reviews}

Photoemission studies of quasiparticle properties 
of a doped hole in the antiferromagnet $\rm Sr_2 CuO_2 Cl_2 $
have revealed\cite{wells,kim}
nearly isotropic dispersion around ${\bf k} = (\pi/2,\pi/2)$,
a bandwidth of around 0.3 eV,
nearly same energy at $(\pi,0)$ and $(0,0)$,
and sharp quasiparticle peaks only near $(\pi/2,\pi/2)$ and $(2\pi/3,0)$.
In the context of photoemission studies,
an overwhelming majority of theoretical studies consider the $t-J$ model,
involving only nearest-neighbour spin coupling
and excluding double occupancy completely, 
thus retaining only the most significant physics of the large $U$ limit.
Quasiparticle renormalization has been obtained
within the self-consistent Born (noncrossing) approximation
involving scattering of mobile carriers due to multiple
emission/absorption of spin waves.
Well defined quasiparticles are obtained near the top of the band at
${\bf k} = (\pi/2,\pi/2)$,
with significantly reduced coherent weight,
and a broad incoherent background.
The experimentally observed energy difference between
${\bf k} = (\pi/2,\pi/2)$ and $(\pi,0)$ states
is too large to be understood within the $t-J$ model.
By including additional next-nearest-neighbour (NNN) 
and NNNN hopping terms $t'$ and $t''$,
which control the $(0,0)-(\pi,0)$ and $(0,0),(\pi,0)-(\pi/2,\pi/2)$
energy separations, respectively, substantially improved fittings have been
obtained,\cite{nazarenko,belinicher,kyung,xiang,eder,laughlin,lee,leung}
with parameter values in the range:
$J=0.12 - 0.15$ eV,
$t=0.3 - 0.4$ eV,
$t'=-(0.07 - 0.14)$ eV,
and $t''=0.05 - 0.09$ eV.
Quasiparticle dispersion fits have also been obtained within
the $t-t'-t''$ Hubbard model at the mean-field level,\cite{karen,duffy}
and within the Born approximation.\cite{chub+morr}
The need for more realistic microscopic models for cuprates
which include additional hopping terms has also been acknowledged
recently from band structure, photoemission, and  
neutron-scattering studies of high-T$_{\rm c}$
and related materials.\cite{nnn1,nnn2,nnn3,nnn4,nnn5,nnn6}
Estimates for $|t'/t|$ range from 0.15 to 0.5.

While electronic excitations in cuprates
can be quantitatively understood within the
$t-J$ model with additional hopping terms,
certain features of the low-energy magnetic excitations
(spin waves) cannot be understood, even qualitatively.
Recent high-resolution inelastic neutron scattering studies
of the spin-wave spectrum of the square-lattice spin-1/2 antiferromagnet
$\rm La_2 Cu O_4 $ have revealed noticeable spin-wave dispersion
along the AF zone boundary.\cite{neutron}
While the simplest explanation for the observed spin-wave dispersion
is in terms of a {\em ferromagnetic} NNN spin coupling $J'$,
it was pointed out\cite{neutron,pallab}
that a more natural explanation lies within the one-band Hubbard model,
for which the strong coupling
expansion in powers of $t/U$ generates extended-range spin couplings.
Fits of the spin-wave spectrum yield $t \approx 0.35$ eV and
$U/t \approx 7$, with $|t'|/t=0.25$.\cite{neutron,pallab}

Given that certain key features of spin-wave excitations in cuprates
have now been ascribed to finite $U$, it would be reassuring if a
unified description of both electronic and magnetic excitations in cuprates
can be obtained with the same set of Hubbard model parameters.
The present work does provide such a unified description.
In this paper we study the quasiparticle properties
of a doped carrier in the AF state of the $t-t'-t''$ Hubbard model
within a mean-field-plus-fluctuation approach.
Specifically, we study the self-consistent fermion self energy
due to fermion-magnon interaction
within the noncrossing (self-consistent Born) approximation.
Not only is a Hubbard model approach to quasiparticle renormalization
evidently more realistic for cuprates, it also provides a complimentary
perspective involving a straightforward physical interpretation.

The perturbative approach for the $t-J$ model
involves a slave-fermion representation to
implement the single-occupancy constraint,
mapping to a holon-spinon Hamiltonian,
and then an expansion in the hopping term $t$.
As the hole hopping is accompanied by upturned or restored spins,
the hole motion
is renormalized due to its coupling with magnon emission and absorption,
and within the self-consistent Born approximation (SCBA),
which is intimately related to the retraceable path approximation,\cite{rice}
the hole self energy amounts to a self-consistent, second-order correction
(in $t$) involving the hole-magnon
scattering.\cite{schmitt-rink,gros,kane,liu,marsiglio,martinez}
Exact diagonalization for small systems has also been carried out.\cite{dagotto}
On the other hand, the fluctuation approach adopted in this paper
involves a many-body expansion in powers of $U$,
wherein the hole renormalization is physically due to
the dynamical transverse fluctuations about the mean field,
and the hole self energy amounts to a self-consistent, second-order
correction (in $U$) involving inelastic, spin-flip scattering
induced by the transverse fluctuations. 
This fluctuation perspective of hole renormalization
is physically relevant for cuprate antiferromagnets,
where the observed Cu moment is reduced by nearly half
due to strong spin fluctuations.
Indeed, this fluctuation-induced quantum correction to the ordered moment
is described by an interband self-energy contribution
involving interband spectral weight transfer,
which is absent in the $t-J$ model approach. 

Reassuringly,
the formally weak-coupling fluctuation approach for the Hubbard model,
when carried to the strong coupling limit,
yields (for the intraband self energy)
exactly the $t-J$ model result of the strong coupling approach,
thus continuously interpolating between the weak and strong coupling limits,
and allowing for quasiparticle studies in the full $U$ range.
In the spin-density-wave (SDW) state for finite $U$, 
several important aspects of correlated electron systems,
which are projected out in the $t-J$ model, can be addressed,
such as interband spectral weight transfer,
one-particle density of states for both bands,
renormalization of the AF band gap,
and the possibility of vanishing band gap even at moderate $U$,
thus providing an understanding of the
antiferromagnetic insulator - paramagnetic metal transition
within a spin/charge fluctuation scenario. 
These finite-$U$ aspects have not been systematically
studied in the full $U$ range in earlier studies.

Earlier studies of self-energy and Green's function corrections
due to spin fluctuations in the AF state of the Hubbard model
have examined one-loop quantum corrections to sublattice
magnetization,\cite{schrieffer,as,quantum,chubukov,chub2,chub3}
ground state energy,\cite{as} and quasiparticle dispersion,\cite{vignale,as}
and later extended to the SCBA level with regard to
quasiparticle dispersion and weight.\cite{sdw-scba,chub+morr}
For intermediate and strong coupling, the one-loop results are
only a poor approximation, as they differ substantially 
from the self-consistent calculations.\cite{sdw-scba}
Only the interband self energy contribution was considered in Ref. [38],
whereas intraband contribution was
argued to be more important for $t/J\sqrt{S} >> 1$
within a $1/S$ expansion in Ref. [14],
where the SCBA self energy was studied in the context of photoemission results.
A finite momentum cutoff in the magnon spectrum was introduced
to model the suppressed weight of zone-boundary magnons,
as indicated by two-magnon Raman scattering.
As the high-energy, zone-boundary magnons contribute significantly
to the SCBA self energy, their elimination biases the quasiparticle
dispersion towards the mean-field result,
which has been noted to yield, for $t'=-0.5J$,
a surprisingly good description of the photoemission
results.\cite{karen,duffy,chub+morr}
Good agreement with experimental data
was obtained for $J/t=0.4$ and $t'=-0.4J$,
with a quasiparticle bandwidth of around $3J$.
Longitudinal and charge fluctuations are negligible in the strong coupling
limit,\cite{schrieffer,chubukov,vignale}
and all fluctuations are small in the weak coupling limit.

Fluctuation self energy has also been examined
in the normal metallic state,
close to the transition to the AF insulator, by assuming
a model dynamical spin susceptibility $\chi(q,\omega)$.\cite{kampf}
In this case, instead of the intra and interband self energies,
one has intra and inter Fermi surface contributions
from scattering states ${\bf k-q}$ lying on same or opposite side of the Fermi surface.
With increasingly long-range AF correlations, the dynamical spin susceptibility
is more sharply peaked at ${\bf q} = {\bf Q}$, the AF wavevector,
and the inter Fermi surface contribution dominates as wavevectors
${\bf k}$ and ${\bf k-Q}$ lie on opposite sides of the Fermi surface.
Formally resembling second-order correction,
the inter Fermi surface self energy shifts the particle (hole) energy
upward (downward), leading to a 
pseudo gap formation in the electronic spectrum with increasing $U$.
The pseudo gap develops into the AF gap as long range spin order sets in,
thus providing a complimentary approach from the paramagnetic side.

Both SCBA studies\cite{sdw-scba,chub+morr}
assume the strong coupling structure of the magnon dispersion
and amplitudes, whereas in the weak coupling limit,
magnon modes with energy approaching the band gap have strongly suppressed
amplitudes.\cite{spectral} The finite-$U$ aspects mentioned above,
such as DOS and AF band gap, have therefore not been systematically studied
in the full $U$ range.

In this paper we examine the finite-$U$ aspects in terms of 
an interplay between classical (mean-field) features of
fermion and magnon dispersion and quantum (self-energy) corrections
on quasiparticle dispersion. The interplay between classical and
quantum effects is manifested in several ways.
For example, the AF state of the Hubbard model at strong coupling involves 
a mean-field dispersion term $4J \gamma_{\bf k}^2$, where $J=4t^2/U$
and $\gamma_{\bf k} = (\cos k_x + \cos k_y)/2$.
Absent in the $t-J$ model, this large classical dispersion term
has an energy scale twice that of the magnon energy,
and therefore quasiparticle renormalizations in the Hubbard and $t-J$ models
are expected to differ with increasing $J$. 
Furthermore, the finite-$U$ induced dispersion term
$4J \gamma_{\bf k}^2$ intrinsically contains effective NNN ($t'=J/2$)
and NNNN ($t''=J/4$) hopping terms,
and these finite-$U$ contributions are entangled with the
physical hopping strengths in the
$t'$ and $t''$ values determined by quasiparticle dispersion fitting
for cuprates within the $t-t'-t''-J$ model.

Another manifestation of the interplay is a competition between
classical and quantum effects. Thus, while quantum correction
lowers the hole energy near $(\pi/2,\pi/2)$ in the $t-J$ and Hubbard models, 
the dispersion term $-4t'\cos k_x \cos k_y$
favours the $(\pi,0)$ state (for positive $t'$).
An effective one dimensional dispersion can thus result 
from a net cancellation of the dispersion along the magnetic zone boundary.

An instance where both classical and quantum effects come together
involves the AF band gap for particle-hole excitations.
Both, the classical dispersion term $-4t'\cos k_x \cos k_y$
and the quantum correction, reduce the band gap significantly,
which may therefore be expected to vanish at a moderate $U$ value.
The antiferromagnetic insulator - paramagnetic metal transition
associated with the vanishing band gap,
relevant for transition-metal oxides such as $\rm V_2 O_3$,
can thus be explored within a spin-fluctuation theory. 

Competition between the two classical dispersion terms in the $t-t'$ Hubbard
model introduces asymmetry in the two AF bands,
in terms of bandwidth and density of states.
The self-energy corrections for added holes and electrons
are therefore expected to be significantly different.
Indeed, we find a dramatically enhanced strong-coupling signature in the 
self-energy correction for the narrow band.

After a brief review of the mean-field AF state
of the $t-t'-t''$ Hubbard model (section II),
the intraband self energy for one added hole (electron)
is obtained in the noncrossing approximation (section III).
Various quasiparticle properties are evaluated for the Hubbard model 
in the full $U$ range (section IV).
The role of finite  $t'$ and $t''$ is studied in section V,
and our conclusions are presented in section VI. 
Some general features associated with the self energy in the AF state
are put in Appendix A, and the interband self energy contribution
is discussed in Appendix B.

\section{Mean-field AF state of the $t-t'-t''$ Hubbard model}
To introduce the notation,
we briefly review the mean-field (Hartree-Fock) state of the
Hubbard model
\begin{equation}
H = 
\sum_{i,\delta,\sigma}  -t_\delta \; 
a_{i, \sigma}^{\dagger} a_{i+\delta, \sigma}
+  U\sum_{i} n_{i \uparrow} n_{i \downarrow} 
\end{equation}
on a square lattice,
with hopping terms $t_\delta = t$, $t'$, and $t''$,
connecting  NN, NNN, and NNNN pairs of sites $(i,i+\delta)$, respectively.
It is convenient to use the two-sublattice basis
as translational symmetry is preserved,
and the Hartree-Fock (HF) Hamiltonian
\begin{equation}
H_0 ^\sigma ({\bf k})=
(\epsilon'_{\bf k} + \epsilon''_{\bf k}) \; {\bf 1} + 
\left [ \begin{array}{cc}
-\sigma \Delta & \epsilon_{\bf k}  \\
\epsilon_{\bf k} & \sigma \Delta  \end{array} \right ]
\end{equation}
for spin $\sigma$, in terms of the free-particle energies
$\epsilon_{\bf k} = -2t(\cos k_x + \cos k_y)$,
$\epsilon'_{\bf k}= -4t'\cos k_x \cos k_y$, and
$\epsilon''_{\bf k} = -2t''(\cos 2k_x + \cos 2k_y)$.
Here $2\Delta = mU$, where $m$ is the sublattice magnetization.

For the classical level (mean-field) fermion propagator,
\begin{equation}
[G^0 _\uparrow ({\bf k},\omega)] =  
\frac
{
\left [ \begin{array}{cc}
\alpha_{\bf k}^2  &  - \alpha_{\bf k} \beta_{\bf k}  \\
- \alpha_{\bf k} \beta_{\bf k} & \beta_{\bf k}^2 \end{array} \right ]
}
{
\omega - E_{\bf k}^{\ominus} -i \eta
}
+
\frac
{
\left [ \begin{array}{cc}
\beta_{\bf k}^2  &  \alpha_{\bf k} \beta_{\bf k}  \\
\alpha_{\bf k} \beta_{\bf k} & \alpha_{\bf k}^2 \end{array} \right ]
}
{
\omega - E_{\bf k}^{\oplus} + i \eta
}
\end{equation}
for spin $\uparrow$,
and $[G^0 _\downarrow ({\bf k},\omega)] =
[\sigma_x] [G^0 _\uparrow ({\bf k},\omega)] [\sigma_x]$
from spin-sublattice symmetry,
where $[\sigma_x]=\left [ \begin{array}{cc}
0 & 1 \\
1 & 0 \end{array} \right ]$.
While the AF band energies
\begin{eqnarray}
E_{\bf k}^{\ominus} &=& (\epsilon' _{\bf k} + \epsilon''_{\bf k})
- (\Delta^2 + \epsilon_{\bf k}^2 )^{1/2} \nonumber \\
E_{\bf k}^{\oplus} &=& (\epsilon' _{\bf k} + \epsilon''_{\bf k})
+ (\Delta^2 + \epsilon_{\bf k}^2 )^{1/2} \; ,
\end{eqnarray}
for the lower and upper bands are modified by $t'$ and $t''$,
the fermion amplitudes
$|{\bf k} \rangle \equiv (\alpha_{\bf k} \; \beta_{\bf k})$ remain unchanged:
\begin{eqnarray}
& & \alpha_{\bf k} ^2 = 
\frac{1}{2}\left ( 1 + \frac{\Delta}
{\sqrt{\Delta^2 + \epsilon_{\bf k} ^2} } \right ) \nonumber \\
& & \beta_{\bf k} ^2 =
\frac{1}{2}\left ( 1 - \frac{\Delta}
{\sqrt{\Delta^2 + \epsilon_{\bf k} ^2} } \right )  \; .
\end{eqnarray}
In the strong coupling limit ($U>>t$),
the majority and minority densities
reduce to $1 - \epsilon_{\bf k}^2 /U^2 $
and $\epsilon_{\bf k}^2 /U^2 $.

Similarly, for the magnon propagator we have 
\begin{equation}
[\chi^{-+}({\bf q},\Omega)] = 
\frac{
\left [ \begin{array}{lr}  
v_{\bf q} ^2 & u_{\bf q} v_{\bf q} \\
u_{\bf q} v_{\bf q} & u_{\bf q} ^2 
\end{array} \right ] }
{\Omega-\Omega_{\bf q}+i\eta}
+
\frac{
\left [ \begin{array}{lr}  
u_{\bf q} ^2 & u_{\bf q} v_{\bf q} \\
u_{\bf q} v_{\bf q} & v_{\bf q} ^2 
\end{array} \right ] }
{\Omega+\Omega_{\bf q}-i\eta} 
\end{equation}
at the classical (ladder-sum) level.
For the Hubbard model in the strong coupling limit
the magnon amplitudes
$|{\bf q} \rangle \equiv (u_{\bf q} \; v_{\bf q})$
and energy $\Omega_{\bf q}$ are given by\cite{as}
\begin{eqnarray}
u_{\bf q}^2 &=& 
\left (1/\sqrt{1-\gamma_{\bf q}^2} + 1 \right )/2
\nonumber \\
v_{\bf q}^2 &=& 
\left (1/\sqrt{1-\gamma_{\bf q}^2} - 1 \right )/2
\nonumber \\
u_{\bf q}v_{\bf q} &=& 
\left (-\gamma_{\bf q}/\sqrt{1-\gamma_{\bf q}^2} \right )/2 \nonumber \\
\Omega_{\bf q} &=& 2J \sqrt{1-\gamma_{\bf q}^2} \nonumber \\
{\rm where} \; \; \gamma_{\bf q} &=& (\cos q_x + \cos q_y)/2 \; .
\end{eqnarray}

\section{Self energy in the noncrossing approximation} 
We now obtain the self energy 
due to fermion-magnon interaction in the noncrossing
(self-consistent Born) approximation (Fig. 1),
in which all noncrossing diagrams are summed,
representing multiple magnon emission and absorption processes.
The magnon lines are considered here at the 
classical (ladder-sum) level.
The first-order quantum corrections to magnon propagator
yield momentum-independent renormalizations of the magnon amplitude
and energy in the strong coupling limit,\cite{quantum}
which can be incorporated easily.

The SCBA self-energy 
\begin{equation}
[\Sigma ({\bf k},\omega)] =
U^2 \sum_{\bf q} \int \frac{d\Omega}{2\pi i}
[\chi^{-+}({\bf q},\Omega)][G_\downarrow ({\bf k-q},\omega - \Omega)] 
\end{equation}
for state $|{\bf k}\uparrow \rangle$ involves
the renormalized propagator for state $|{\bf k-q}\downarrow \rangle$.
However, from spin-sublattice symmetry,
the fermion propagators are identical if their spin and sublattice indicies 
are simultaneously switched, so that
\begin{eqnarray}
[G_\downarrow({\bf k-q},\omega-\Omega)] &=& 
[\sigma_x][G_\uparrow({\bf k-q},\omega-\Omega)][\sigma_x] \nonumber \\
&=& 
\frac{
[\sigma_x]|{\bf k-q} \rangle \langle {\bf k-q} | [\sigma_x] 
}
{\omega-\Omega -E_{\bf k-q}^0 - \Sigma_{\bf k-q}(\omega-\Omega)}
\; , \nonumber  \\
\end{eqnarray}
using Eq. (A1).
The above replacement in Eq. (8) thus yields a self-consistent equation
for the fermion self energy $\Sigma_{\bf k}(\omega)$
which is independent of spin.
In the following, we consider only the dominant
intraband self energy contribution
in which the intermediate fermion state $|{\bf k-q}\downarrow \rangle$
lies in the same band as the state $|{\bf k}\uparrow \rangle$.
The interband contribution is discussed in Appendix B.
Due to an exact cancellation with the Hartree self energy,
the interband self energy does not contribute to gap renormalization
in the strong coupling limit, and the quasiparticle coherent weight
reduction is small, even in two dimensions.

%fig1
\begin{figure}
%\vspace*{10mm}
\hspace*{-7mm}
\epsfig{figure=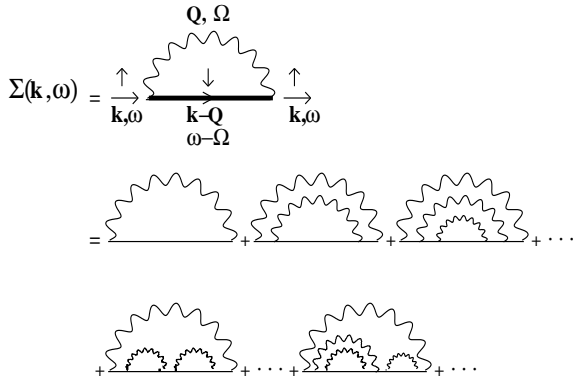,width=75mm,height=50mm}
\vspace{0mm}
\caption{Self energy in the noncrossing approximation.
Wavy lines represent the magnon propagator.
The bare fermion-magnon interaction vertex is $U$.}
\end{figure}

\subsection{One hole (electron) in the AF}
The situation is simplest for a single hole introduced in a filled
lower AF band,
as the available scattering states $|{\bf k-q}\rangle$ 
span all of the lower AF band. 
From particle-hole symmetry, the situation is symmetrical
for an added electron to the empty upper band,
under sign change of $t'$ and $t''$. 
For finite doping,
there are two intraband self energy contributions arising from 
scattering states $|{\bf k-q}\rangle$ above or below the Fermi energy,
involving retarded or advanced magnon poles in Eq. (8)
The local reduction in magnetic order near the hole is due to
intraband spectral weight transfer across the Fermi energy. 

The motion of a hole in an antiferromagnetic background 
has been studied extensively in the recent past.\cite{reviews}
%\cite{brinkman,gros,schmitt-rink,kane,shraiman,sachdev,trugman,dagotto,
%szczepanski,dagotto2,liu1,liu2,marsiglio,martinez,} 
The string of upturned spins and broken antiferromagnetic bonds left in its wake
renders the hole motion highly incoherent,
and the description of the hole carrying along 
the locally scrambled AF spin arrangement  
is a highly nontrivial many-body theoretical problem. 

Several aspects of the hole propagator,
such as the spectral function, quasiparticle dispersion, bandwidth and weight,
have been studied for the $t-J$ model within the noncrossing
(self-consistent Born) 
approximation.\cite{schmitt-rink,gros,kane,liu,marsiglio,martinez}
Comparison with exact diagonalization results\cite{dagotto} show the      
SCBA results to be in good agreement for small $J$,
and contributions of crossing diagrams have been argued
to be small.\cite{liu,chub+morr}

As $G_\downarrow ({\bf k-q},\omega - \Omega)$ in Eq. (8)
is an advanced propagator (for the intraband contribution),
representing scattering states for the added hole, 
the $\Omega$ integral (replaced by a contour integral over the upper half plane)
picks a contribution only from the advanced part of the magnon propagator
in Eq. (6).
It is convenient to introduce a composite amplitude
\begin{equation}
|\sigma({\bf k},{\bf q})\rangle = |{\bf q} \rangle \times 
[\sigma_x]|{\bf k-q} \rangle 
= \left (\begin{array}{c} u_{\bf q}\; \beta_{\bf k-q} \\ 
v_{\bf q}\; \alpha_{\bf k-q} \end{array} \right ) \; ,
\end{equation}
for the magnon-hole interaction vertex,
in terms of which we obtain for the hole self energy
\begin{eqnarray}
\Sigma_{\bf k}(\omega) &\equiv & 
\langle {\bf k} | [\Sigma({\bf k},\omega)] | {\bf k} \rangle 
=(\alpha_{\bf k}^* \;\; \beta_{\bf k}^*) 
\left [ \Sigma({\bf k},\omega) \right ]
\left (
\begin{array}{c}
\alpha_{\bf k} \\ \beta_{\bf k}
\end{array} \right ) \nonumber \\
&=& U^2 \sum_{\bf q} 
\frac{\langle {\bf k} |\sigma({\bf k},{\bf q})\rangle 
\langle \sigma({\bf k},{\bf q}) | {\bf k} \rangle} 
{\omega+ \Omega_{\bf q} -E^\ominus_{\bf k-q} - 
\Sigma_{\bf k-q} (\omega+\Omega_{\bf q}) } \nonumber \\
&=&
U^2 \sum_{\bf q} 
\frac{ 
(\alpha_{\bf k} u_{\bf q}\; \beta_{\bf k-q} + 
\beta_{\bf k} v_{\bf q}\; \alpha_{\bf k-q} )^2 }
{\omega+ \Omega_{\bf q} -E^\ominus_{\bf k-q} - 
\Sigma_{\bf k-q} (\omega+\Omega_{\bf q}) } 
\; . \nonumber \\
\end{eqnarray}

The above yields a self-consistent equation for the hole self energy 
in the form of an integral equation,
which describes the hole renormalization 
due to multiple magnon emission and absorption processes
as it moves in the AF background,
yielding a quantum correction to the classical (mean-field) hole dispersion.
While this intraband self energy significantly redistributes spectral
weight in the hole (particle) band, leading to bandgap renormalization,
it leaves the integrated spectral weight unchanged,
so that the spin densities and ordered moment are unaffected.
Within the noncrossing approximation, Eq. (11) is valid for arbitrary $U$,
in the whole range from weak to strong coupling,
and for extended-range hopping terms.
As the self energy depends on the classical dispersion,
explicitly through  $E^\ominus_{\bf k-q}$
and implicitly through the magnon energy $\Omega_{\bf q}$,
an interesting interplay is expected 
between classical and quantum effects on quasiparticle energies.

\section{Self energy for the Hubbard model}
\subsection{Strong coupling limit}
We now consider the self energy for the Hubbard model in the
analytically simple strong coupling limit $(U/t \rightarrow \infty)$,
and compare with the $t-J$ model result.
We write the classical (mean-field) hole energy 
\begin{equation}
E_{\bf k}^\ominus = -\sqrt{\Delta^2 + \epsilon_{\bf k}^2 }
\approx -\Delta - 4J_c\gamma_{\bf k}^2 
\end{equation}
where a classical energy scale $J_c=4t^2/U$ has been introduced 
to distinguish the classical dispersion term $4J_c\gamma_{\bf k}^2$
generated at strong coupling.
Substituting in Eq. (11) the fermion and magnon amplitudes and energies
from Eqs. (5), (7), and (12),
and shifting the energy by $\Delta$ 
to bring the zero of the energy scale to the lower band edge,
the self energy is compactly written as 
\begin{equation}
\Sigma_{\bf k}(\omega) 
= t^2 z^2 
\sum_{\bf q} 
\frac{(u_{\bf q} \gamma_{\bf k-q} + v_{\bf q} \gamma_{\bf k} )^2 }
{\omega+\Omega_{\bf q}+4J_c \gamma_{\bf k-q}^2  
- \Sigma_{\bf k-q} (\omega+\Omega_{\bf q}) } \; ,
\end{equation}
which is exactly the SCBA result for the $t-J$ model,\cite{schmitt-rink,gros,kane}
apart from a classical dispersion term absent in the $t-J$ model.
This shows that the formally weak-coupling fluctuation approach,
involving a many-body expansion in powers of $U$,
when carried to the strong coupling limit,
yields the strong-coupling result for the hole self energy as well.
Similar exact correspondence
has been noted earlier for magnon dispersion,
both at the classical level and including quantum corrections,
sublattice magnetization,
ground-state energy,
two-magnon Raman scattering etc. 

While the effective fermion-magnon scattering matrix element is reduced
from O$(U)$ to O($t$) due to the minority fermion amplitude $\beta$,
the gapless energy denominator in this intraband term
leads to a self-consistent self energy of O($t$),
which is much greater than the residual interband contribution of O($J$),
surviving the leading-order cancellation (see Appendix B).

\subsection{Strong-coupling results}
The self-consistent numerical evaluation of the self energy  
was carried out on a $52\times 52$ grid in ${\bf k}$ space, 
and a frequency interval $\Delta \omega =0.1$ 
for $\omega$ in the range $-10 < \omega < 10$.
The self energy was iteratively evaluated, 
starting with $\Sigma_{\bf k}(\omega) = 0$.
Typically, self-consistency was achieved within ten iterations.
For $J_c=0$, our numerical results are in exact agreement with 
the $t-J$ model results of Liu and Manousakis.\cite{liu}

%2
\begin{figure}
\vspace*{0mm}
\hspace*{-7mm}
\epsfig{figure=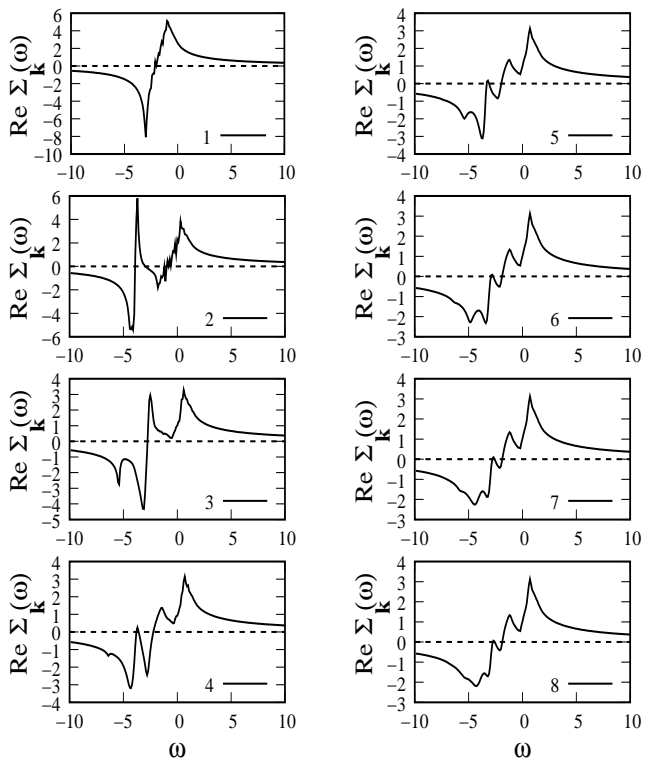,width=85mm,height=100mm}
\vspace{0mm}
\caption{Variation of self energy with iterations, effectively
illustrating the role of successively higher-order magnon processes.
Here $J=0.5$ and ${\bf k}=(\pi/2,\pi/2)$.}
\end{figure}

%\begin{figure}
%\vspace*{5mm}
%\hspace*{-5mm}
%\epsfig{figure=Resig.eps,width=85mm,height=80mm}
%\vspace{0mm}
%\caption{Comparison of the first-order and self-consistent results
%$for ${\rm Re} \; \Sigma_{\bf k} (\omega)$.}
%\end{figure}

%3
\begin{figure}
\vspace*{-10mm}
\hspace*{-5mm}
\epsfig{figure=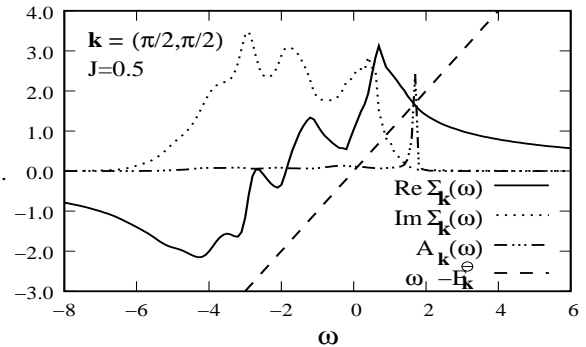,width=77mm,height=60mm}
\vspace{0mm}
\caption{The hole spectral function $A_{\bf k} (\omega)$ shows the
quasiparticle peak at the intersection of $\omega-E_{\bf k}^\ominus $ with
${\rm Re} \; \Sigma_{\bf k} (\omega)$.}   
\end{figure}

%\begin{figure}
%\vspace*{5mm}
%\hspace*{-7mm}
%\epsfig{figure=kcomp.eps,width=80mm,height=50mm}
%\vspace{0mm}
%\caption{Comparison of the quasiparticle energies for ${\bf k}=(\frac{\pi}{2},\frac
%{\pi}{2})$ and $(\frac{\pi}{2},0)$.}
%\end{figure}

The hole self-energy variation with successive iterations (Fig. 2)
effectively illustrates the role of multiple magnon emission and
absorption processes.
The first iteration yields the single magnon contribution, 
and successive iterations incorporate higher-order magnon processes.
The significant difference, especially for $J<<t$, 
between the first-order and the self-consistent results,
indicates the relative importance of multi-magnon processes.

From Eq. (A1),
intersection of $\omega-E_{\bf k}^\ominus$ with
${\rm Re} \; \Sigma_{\bf k} (\omega)$ yields the
renormalized quasiparticle energy $E_{\bf k} ^*$
where the spectral function peaks, as shown in Fig. 3
for ${\bf k}=(\frac{\pi}{2},\frac{\pi}{2})$.
The quadratic growth of ${\rm Im} \; \Sigma_{\bf k} (\omega)$
near $\omega \approx E_{\bf k} ^*$ is due to the long-wavelength magnon modes.
In Eq. (13), neglecting the weak $q^2$ dependence of
$\gamma_{\bf k-q}$ and $\Sigma_{\bf k-q}$ in comparison with the linear
dependence $\Omega_{\bf q} = \sqrt{2}Jq$ of the magnon energy
for small $q$, one obtains
\begin{eqnarray}
{\rm Im} \; \Sigma_{\bf k} (\omega) &\sim & t^2 z^2
\int q dq (1/q) (q_x + q_y )^2 \delta
(\omega + \Omega_{\bf q} + {\rm Re} \; \Sigma_{\bf k} )
\nonumber \\
&\sim & (\omega + E_{\bf k} ^*)^2 \; .
\end{eqnarray}
Magnon damping at finite doping due to intraband particle-hole excitations
is therefore expected to be important for quasiparticle damping.

%4
\begin{figure}
\vspace*{-70mm}
\hspace*{-38mm}
\psfig{figure=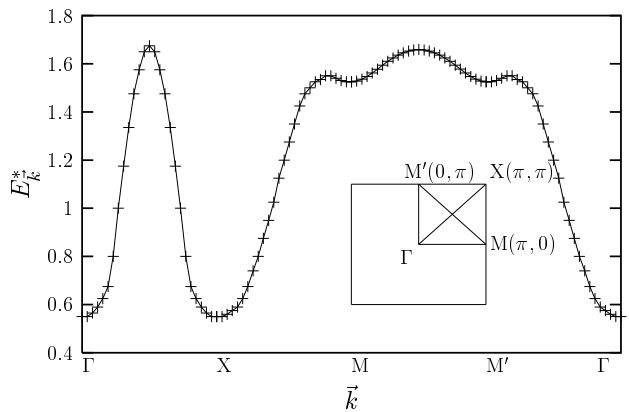,width=140mm}
\vspace{-77mm}
\caption{Quasiparticle dispersion along different symmetry directions, for $J=0.5$.}
\end{figure}

%5
\begin{figure}
\vspace*{-70mm}
\hspace*{-38mm}
\psfig{figure=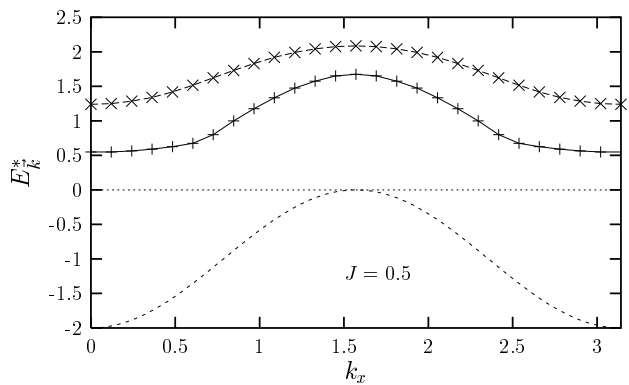,width=140mm}
\vspace{-78mm}
\caption{Quasiparticle dispersion along the $\rm \Gamma-X$ direction
with ($+$) and without ($\times$) the classical dispersion term $J_c$,
along with the mean-field dispersion.}
\end{figure}

The large classical term $4J_c \gamma_{\bf k}^2 $
suppresses the self-energy correction in Eq. (13),
in addition to providing dispersion at the mean-field level,
and these combined effects of $J_c$ on quasiparticle dispersion
are discussed below. Figure 4 shows the quasiparticle dispersion
along different symmetry directions for $J=0.5$.
Here the quasiparticle energies are obtained from the
peak in the spectral function $A_{\bf k}(\omega)$
occuring at lowest hole energy.
The small dispersion developed along the $\rm M-M'$ direction
(magnetic zone boundary) is purely quantum mechanical.
The substantially different curvatures at ${\bf k}=(\pi/2,\pi/2)$
along the $\rm \Gamma-X$ and $\rm M-M'$ directions
shows a highly anisotropic quasiparticle dispersion with
different effective masses.
The quasiparticle energy is maximum for ${\bf k}=(\pi/2,\pi/2)$.

%6
\begin{figure}
\vspace*{-70mm}
\hspace*{-38mm}
\psfig{figure=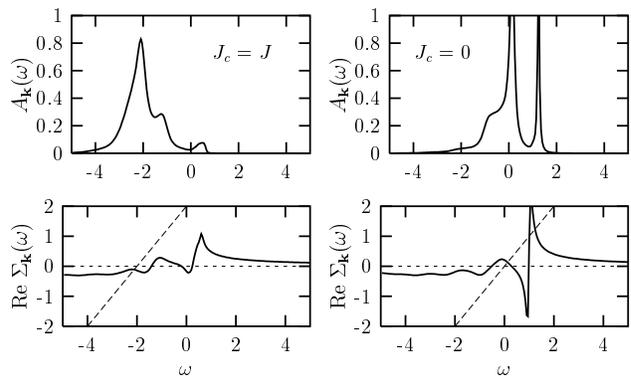,width=140mm}
\vspace{-78mm}
\caption{The ${\bf k}=(0,0)$ spectral function is significantly modified
by the finite-$U$ induced classical dispersion term $J_c$,
as shown in terms of the intersection of
${\rm Re} \; \Sigma_{\bf k} (\omega)$ with the line
$\omega + 4J_c \gamma_{\bf k}^2$. Here $J=0.5$.}
\end{figure}

%7
\begin{figure}
\vspace*{5mm}
\hspace*{-5mm}
\epsfig{figure=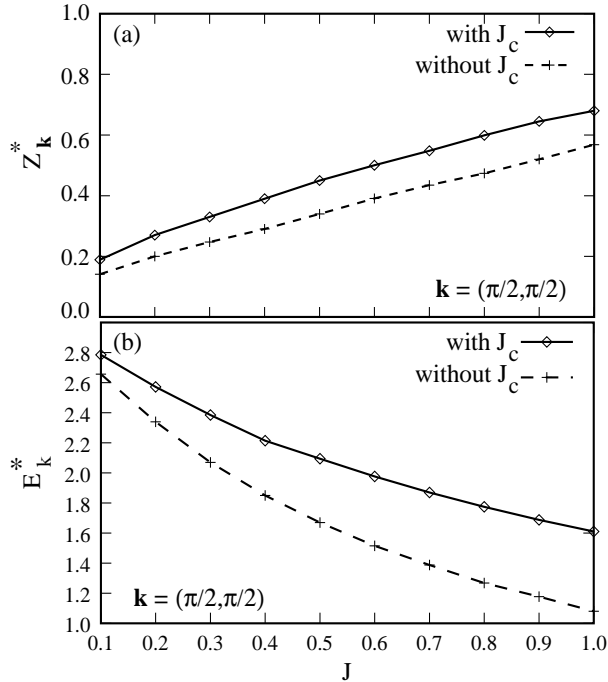,width=80mm,height=90mm}
\vspace{0mm}
\caption{Comparison of quasiparticle weight $Z_{\bf k}$ and
energy $E_{\bf k} ^*$ with and without $J_c$,
showing increasing differences with $J$.}
\end{figure}

Figure 5 shows a comparison of the quasiparticle dispersion
along the $\rm \Gamma-X$ direction with and without $J_c$,
providing a comparison with the SCBA results for the $t-J$ model 
$(J_c =0)$.
The classical dispersion term $J_c$
reduces the quasiparticle energy renormalization,
and significantly flattens the
dispersion near the $\rm \Gamma$ and $\rm X$ points,
and also increases the quasiparticle bandwidth. 
Furthermore, for ${\bf k}=(0,0)$, 
the spectral weight of the highest-energy peak (lowest-energy for hole)
is drastically reduced by $J_c$, in sharp contrast with the $J_c=0$ case,
as shown in Fig. 6.
This finite-$U$ induced feature is quite significant
with regard to photoemission studies of cuprates,
where the ${\bf k}=(0,0)$ peak is seen to be extremely weak
in comparison with $(\pi/2,\pi/2)$.

%8
\begin{figure}
\vspace*{-70mm}
\hspace*{-38mm}
\psfig{figure=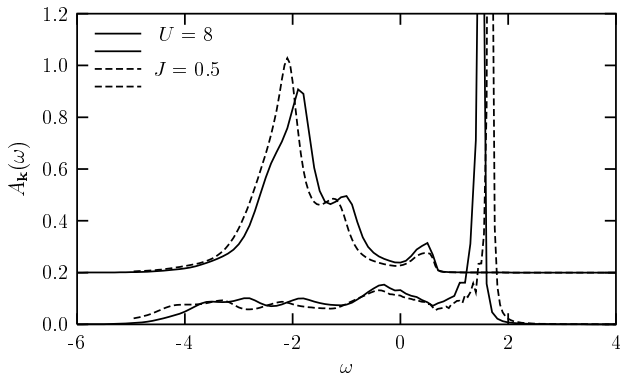,width=140mm}
\vspace{-78mm}
\caption{Comparison of strong-coupling and finite-$U$ spectral functions
for ${\bf k}=(\pi/2,\pi/2)$ and $(0,0)$.}
\end{figure}

%9
\begin{figure}
\vspace*{5mm}
\hspace*{-5mm}
\epsfig{figure=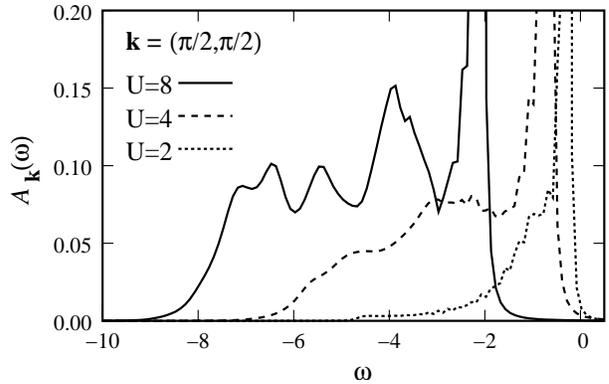,width=80mm,height=50mm}
\vspace{0mm}
\caption{Increasingly incoherent hole spectral function
with increasing $U$, for ${\bf k}=(\pi/2,\pi/2)$.}
\end{figure}

%10
\begin{figure}
\vspace*{5mm}
\hspace*{-5mm}
\epsfig{figure=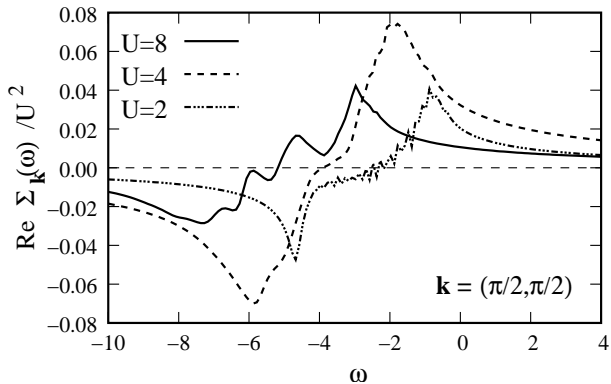,width=80mm,height=50mm}
\vspace{0mm}
\caption{Approximate $U^2$ scaling of the real part of self energy,
showing a dominant single-magnon contribution with decreasing $U$.}
\end{figure}

%11
\begin{figure}
\vspace*{-70mm}
\hspace*{-38mm}
\psfig{figure=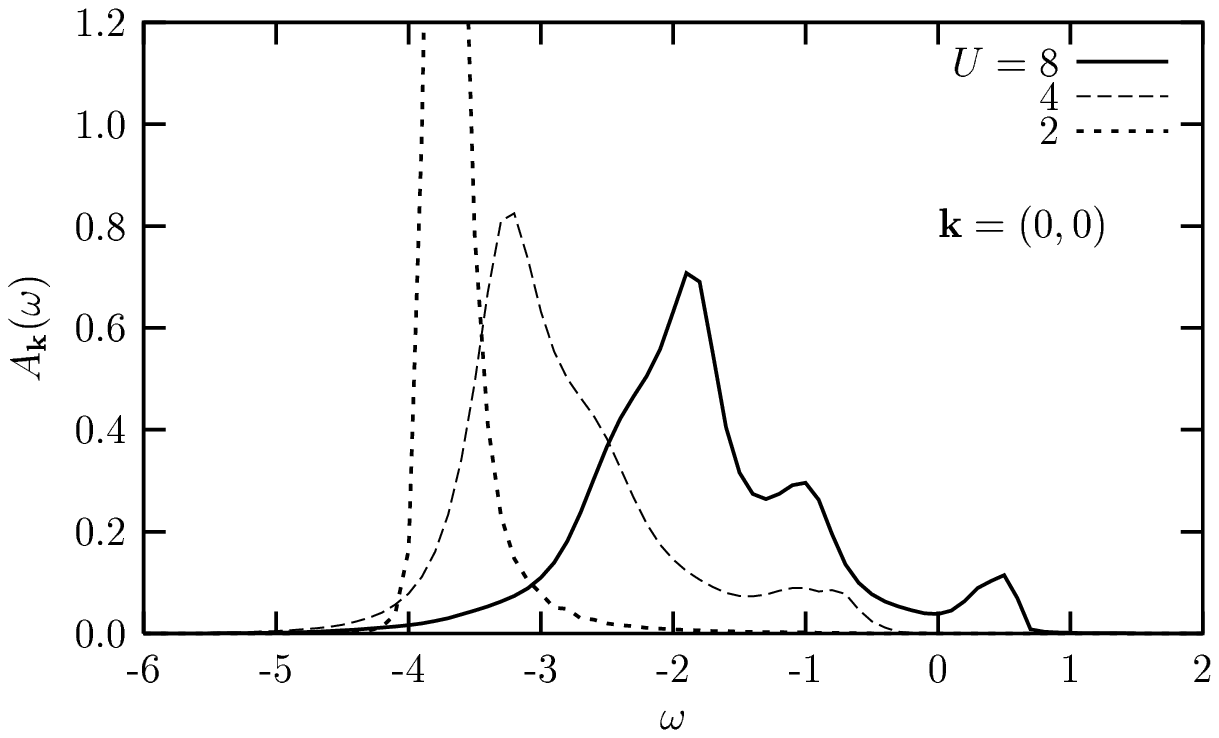,width=140mm}
\vspace{-78mm}
\caption{Evolution of the ${\bf k}=(0,0)$ spectral function 
with decreasing $U$.}
\end{figure}

The enhancement in the self-energy correction
with decreasing $J$ is seen in Fig. 7,  
both through (a) decreasing quasiparticle weight for ${\bf k}=(\pi/2,\pi/2)$
signifying enhancement in incoherent part due to multi-magnon processes,
and (b) increasing quasiparticle energy renormalization.
Comparison with the $t-J$ model results ($J_c = 0$)
illustrates the increasing suppression of quantum correction
by the classical dispersion term with $J$.       

\subsection{Intermediate and weak coupling}
Self energy for finite $U$ is of special interest
as several additional features of the AF state can be explored.
These include one-particle density of states for both bands,
renormalized band gap, and
the possibility of vanishing band gap and a metal-insulator transition
driven by spin fluctuations.
Furthermore, the magnon nature is qualitatively modified at finite $U$,
as studied earlier in detail in the 
intermediate,\cite{inter} and weak\cite{weak} coupling limits.
Extended-range spin couplings generated at intermediate $U$
remove the degeneracy in the magnon spectrum along the AF zone boundary.
Also, the decreasing AF band gap $2\Delta$ begins to play a significant role,
and for small $U$ the magnon amplitude is strongly suppressed
for modes with energy approaching $2\Delta$.\cite{spectral}
These features were not explored in earlier SDW-SCBA
calculations\cite{sdw-scba,chub+morr} for the Hubbard model,
where only strong coupling expressions for magnon
amplitude and energy were employed,
and are naturally beyond the scope of the $t-J$ model
in which the upper AF band is projected out,
and the magnon dispersion corresponds to NN spin coupling only.

For arbitrary $U$,
the hole self energy $\Sigma_{\bf k} (\omega)$ is determined from Eq. (11),
with classical fermion amplitudes and energies taken from Eqs. (4) and (5),
and the classical (ladder-sum) magnon energy $\Omega_{\bf q}$
and amplitudes $u_{\bf q}$, $v_{\bf q}$ determined numerically,
as described earlier.\cite{inter,weak}

%10
%\begin{figure}
%\vspace*{5mm}
%\hspace*{-5mm}
%\epsfig{figure=del.eps,width=80mm,height=50mm}
%\vspace{0mm}
%\caption{The renormalised band gap $E_g ^*$ is nearly half the mean-field band gap
%$E_g ^0$ at intermediate coupling, and approaches $E_g ^0$ with decreasing $U$.}
%\end{figure}

%12
\begin{figure}
\vspace*{-70mm}
\hspace*{-38mm}
\psfig{figure=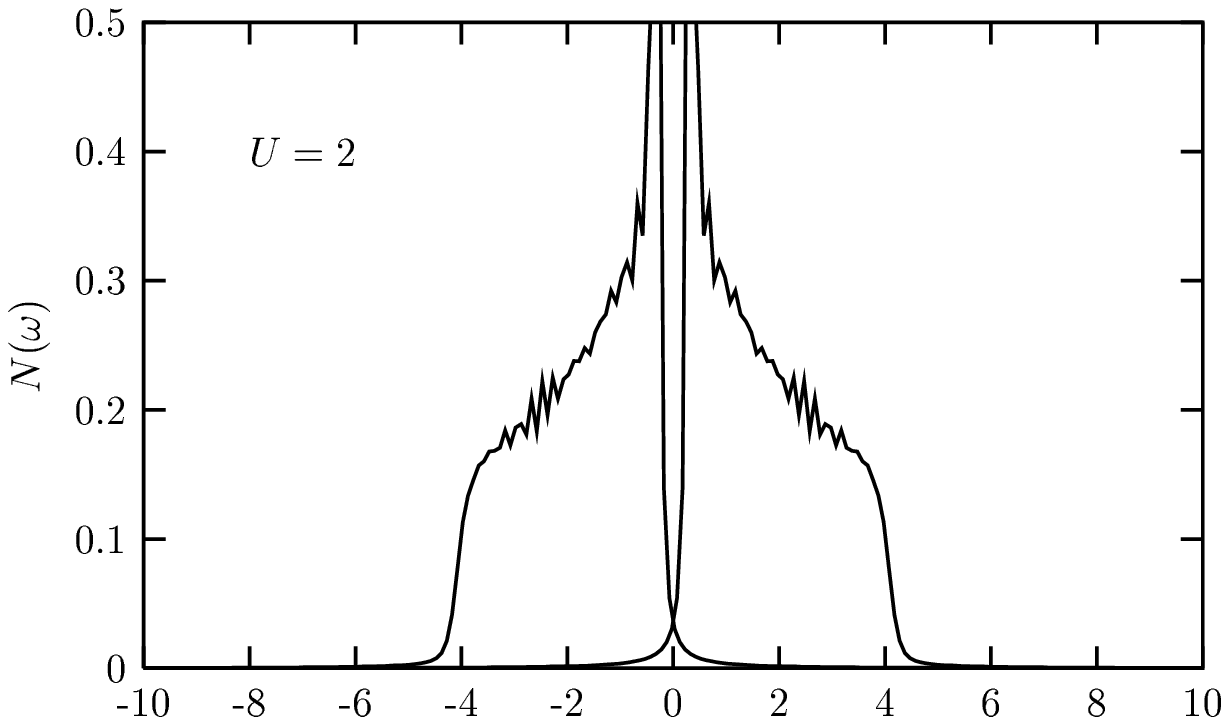,width=140mm}
\vspace{-75mm}
%\caption{.}
%\end{figure}
%\begin{figure}
\vspace*{-77mm}
\hspace*{-38mm}
\psfig{figure=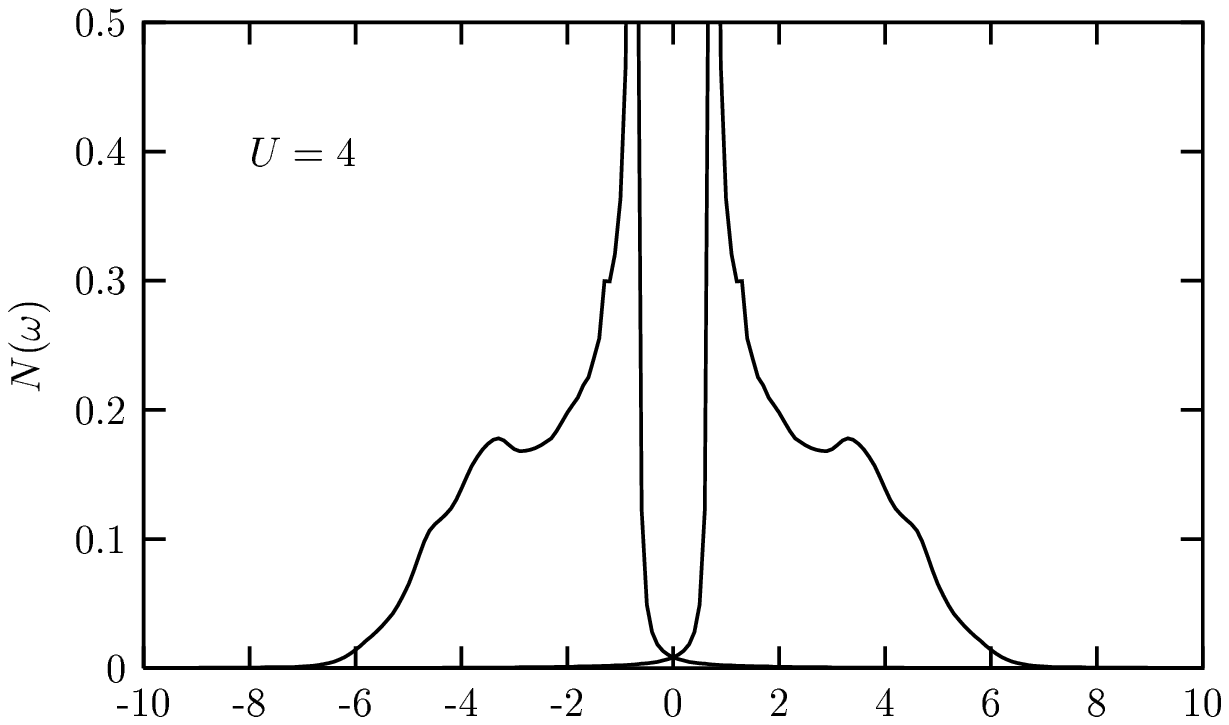,width=140mm}
\vspace{-75mm}
%\caption{.}
%\end{figure}
%\begin{figure}
\vspace*{-77mm}
\hspace*{-38mm}
\psfig{figure=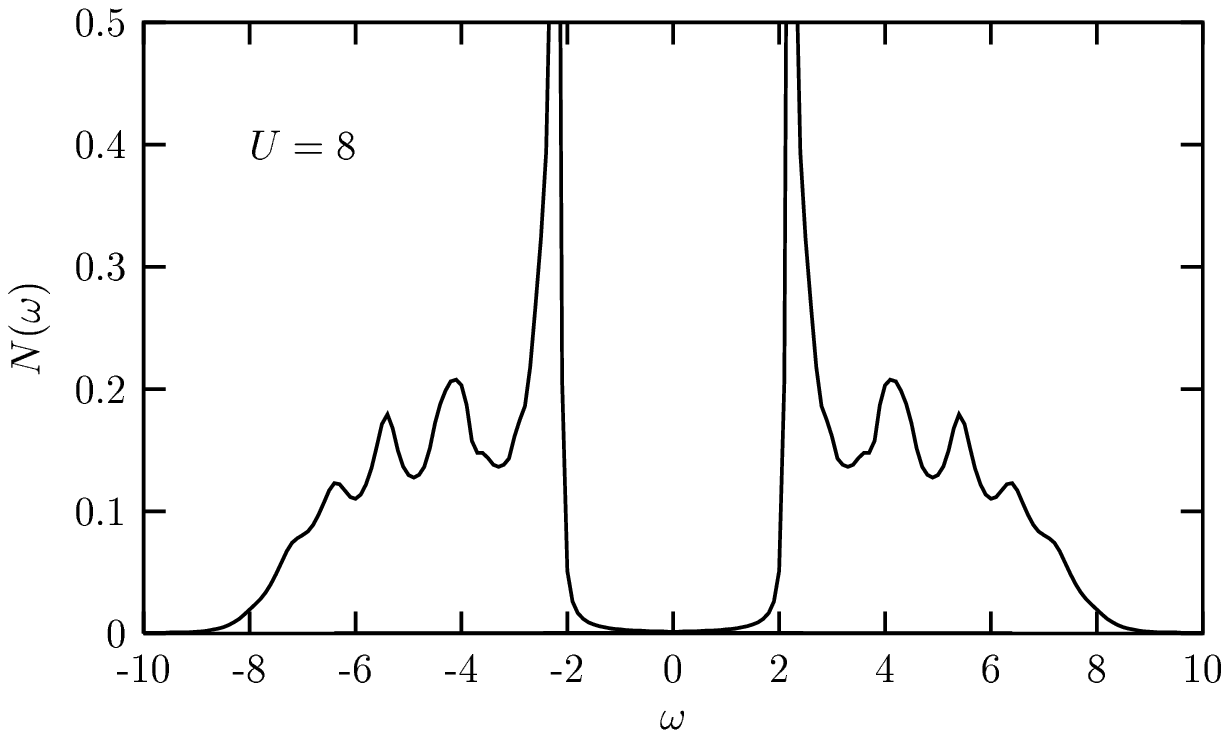,width=140mm}
\vspace{-80mm}
\caption{Renormalized one-particle (tunneling) density of states for $U=8,4,2$.
The corresponding mean-field gaps are $2\Delta = 7.2,2.8,0.75$.}
\end{figure}

%
%\begin{figure}
%\vspace*{5mm}
%\hspace*{-7mm}
%\epsfig{figure=dosu-t.eps,width=80mm,height=130mm}
%\vspace{0mm}
%\caption{DOS for different size of lattice (a)$N=8$, (b) $N=10$,
%(c) ${\rm N}=50{times}50$
%as the size of lattice increases come close.For small lattice DOS does not have much %structure.As the size of lattice
%increases perhaps the other states of large spectral weight apper which increase the %structure.}
%\end{figure}

Comparison of spectral functions evaluated for
intermediate $U$ with strong coupling results (Fig. 8)
shows a noticeable reduction in the quasiparticle bandwidth.
With decreasing $U$, 
the hole spectral function becomes increasingly coherent (Fig. 9),
the self energy approximately scales as $U^2$ (Fig. 10),
and the shape of ${\rm Re}\Sigma_{\bf k}(\omega)$ 
increasingly resembles the single-magnon result (Fig. 3),
indicating dominant lowest-order ($U^2$) contribution,
as expected.
Evolution of the ${\bf k}=(0,0)$ spectral function (Fig. 11) shows that
the peak at lowest hole energy remains nearly incoherent down to very low $U$ values.

The one-particle (tunneling) density of states
is shown in Fig. 12 for $U=8,4,2$
(corresponding to mean-field gaps $2\Delta = 7.2,2.8,0.75$),
displaying the smooth approach
towards the free-particle DOS with decreasing $U$.
The band gap remains finite no matter how small $U$ is. 
States are pulled within the classical energy gap $2\Delta$ 
due to the fermion-magnon scattering,
and the resulting fermion spectral function incoherence 
spreads the DOS over a broader frequency range.
Pseudo-gap feature appears in the strong coupling limit.
These DOS features are in good agreement with
exact diagonalization results.\cite{dagotto}

For all $U$, the state ${\bf k}=(\pi/2,\pi/2)$
continues to be the lowest-energy hole state,
and therefore, from particle-hole symmetry,
also the lowest-energy electron state in the upper band. 
The minimum electron-hole pair excitation energy
therefore yields the renormalized band gap
$E_{\rm g} = 2|E_{\bf k} ^* |$, with ${\bf k} = (\pi/2,\pi/2)$.
We find that the ratio $E_{\rm g} ^*/E_{\rm g}^0$ is reduced to nearly 1/2
in a broad $U$ range, as shown in Fig. 13,
and approaches 1 in both weak and strong coupling limits,
although with fundamentally different behaviour.
The weak-coupling trend indicates that the band gap remains finite for all $U$.
For $U \ge 8$,
we have used the strong-coupling self energy from Eq. (13).

%13
\begin{figure}
\vspace*{5mm}
\hspace*{-5mm}
\epsfig{figure=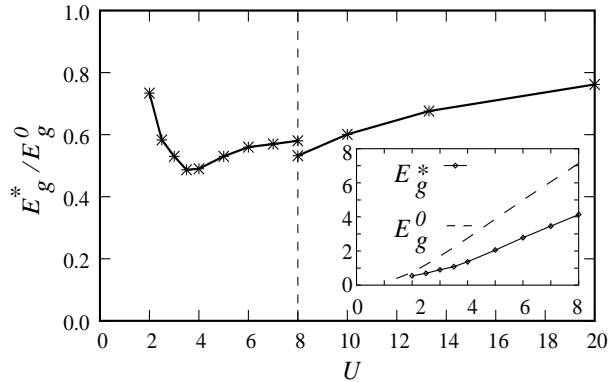,width=80mm,height=50mm}
\vspace{0mm}
%\vspace*{-32mm}
%\hspace*{28mm}
%\epsfig{figure=del.eps,width=40mm,height=25mm}
%\vspace{5mm}
\caption{The ratio $E_g ^*/E_g ^0$ of renormalized and classical band gaps
is nearly 1/2 for intermediate coupling,
and approaches 1 in both weak and strong coupling limits.}
\end{figure}

\section{Role of hopping terms $t',t''$}
Finite hopping terms $t',t''$ add an additional dimension
to the competition between classical and quantum effects on
quasiparticle dispersion.
Thus, the classical dispersion term $\epsilon_{\bf k}^{'}$ introduces a
dispersion along the MM' direction of the Brillouin zone,
renders the two AF bands asymmetrical,
yielding narrow and wide bands for a given $U$ value,
and also significantly reduces the band gap,
making it possible for quantum fluctuation effects
to close the gap at a moderate $U$ value.
Furthermore, the frustration due to competing AF spin couplings
generated by $t',t''$ leads to magnon softening,\cite{soft}
and the reduced magnon energy scale 
is expected to enhance the self-energy correction.
This frustration effect is absent within the $t-t'-t''-J$ model
having only NN AF spin coupling.

%14
%\begin{figure}
%\vspace*{5mm}
%\hspace*{-5mm}
%\epsfig{figure=tpr.eps,width=80mm,height=50mm}
%\vspace{0mm}
%\caption{Variation of the quasiparticle energy difference 
%between the $(\pi/2,\pi/2)$ and $(\pi,0)$ states.}
%\end{figure}

%14
\begin{figure}
\vspace*{-70mm}
\hspace*{-38mm}
\psfig{figure=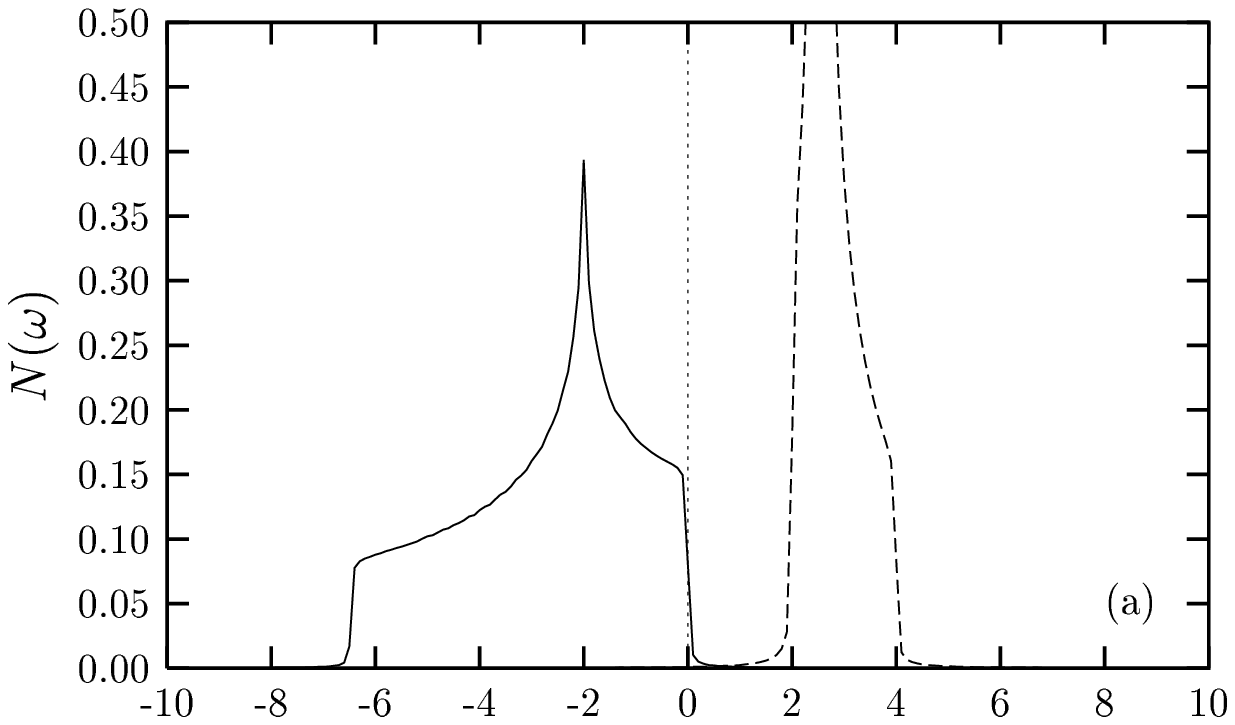,width=140mm}
\vspace{-75mm}
%\caption{.}
\vspace*{-77mm}
\hspace*{-38mm}
\psfig{figure=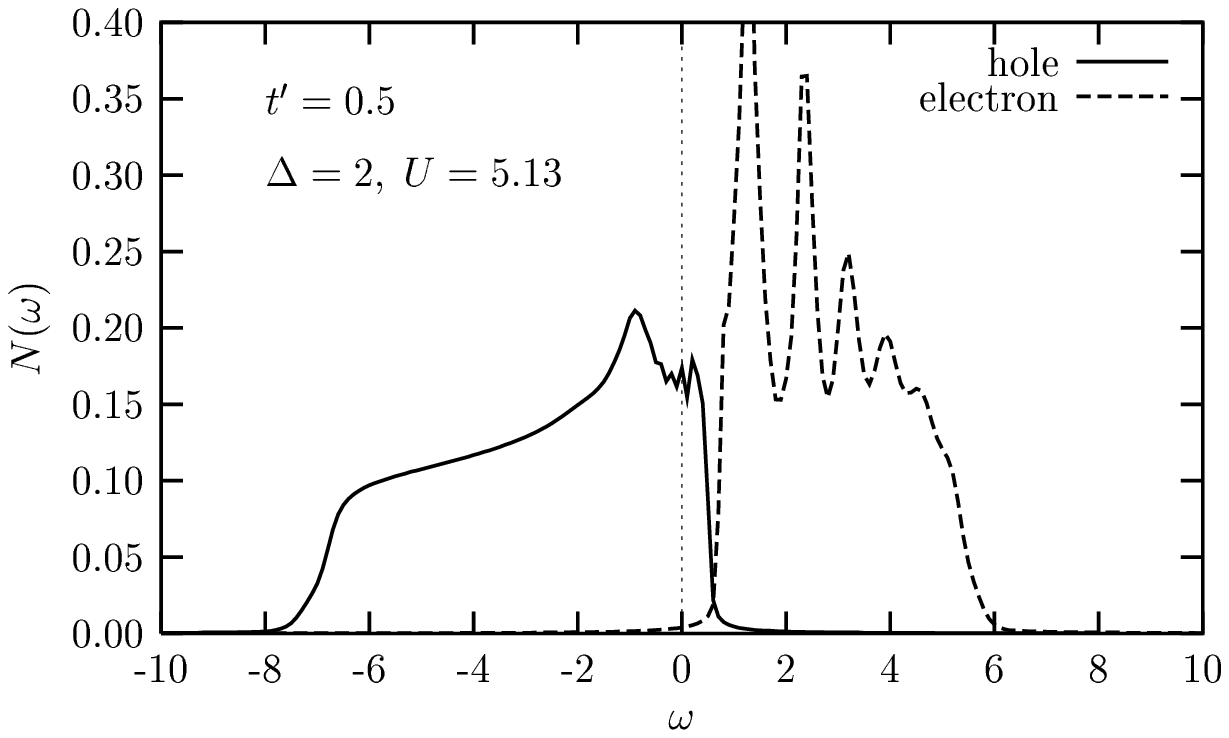,width=140mm}
\vspace{-80mm}
\caption{Classical (a) and renormalized (b)
one-particle density of states for one added hole and electron,
showing the vanishing of the energy gap.
The classical-level band asymmetry strongly
influences the quantum corrections.}
\end{figure}

We have evaluated the hole self energy for finite $U$
and positive $t'$ using Eq. (11), with appropriate classical
fermion energies and amplitudes given in Eqs. (4) and (5).
The quasiparticle energy difference
between the $(\pi/2,\pi/2)$ and $(\pi,0)$ states
decreases nearly linearly with $t'$, and vanishes for $t' \approx 0.1$,
resulting in a nearly one dimensional quasiparticle dispersion
along the $\rm \Gamma-X$ direction.
For higher $t'$ values, the classical effect dominates,
and ${\bf k}=(\pi,0)$ is the lowest-energy hole state.
From particle-hole symmetry,
these results also apply to an added electron in the upper band
for negative $t'$. 

Figure 14 shows the one-particle density of states
for $t'=0.5$ and $U \approx 5$.
While a substantial classical gap remains,
the renormalized band gap is seen to just vanish.
However, as the lowest-energy hole and electron states correspond
to different momenta ---  $(\pi,0)$ for hole and $(\pi/2,\pi/2)$
for electron --- the band gap is indirect,
yielding a finite optical gap.
A further reduction in $U$ will lead to gapless
particle-hole excitations with same momentum.

For one added electron to the narrow upper band,
the characteristic strong-coupling signature of multi-magnon processes
and the large energy shift relative to the mean-field DOS
confirms the strong fermion-magnon scattering associated with a narrow band.
The pseudo-gap feature is also significantly enhanced.
On the other hand, for one added hole to the wide lower band,
significantly weaker fermion-magnon scattering is evident.

%15
\begin{figure}
\vspace*{-70mm}
\hspace*{-38mm}
\psfig{figure=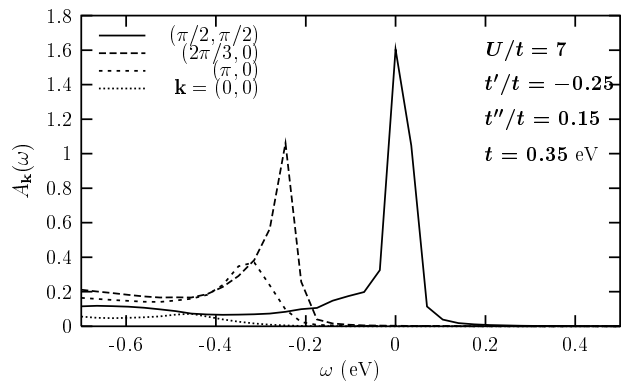,width=140mm}
\vspace{-78mm}
\caption{Spectral function shows sharp quasiparticle
peaks for ${\bf k}=(\pi/2,\pi/2)$ and
$(2\pi/3,0)$, as seen in photoemission experiments on cuprates.}
\end{figure}

%16
\begin{figure}
\vspace*{5mm}
\hspace*{-5mm}
\epsfig{figure=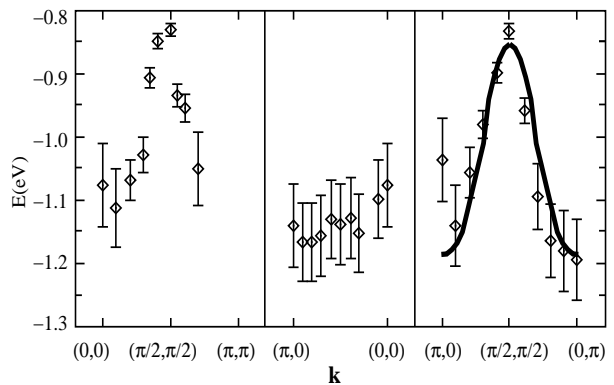,width=80mm,height=50mm}
\vspace{0mm}
\caption{Quasiparticle dispersion along $(\pi,0)$ --- $(0,\pi)$
with same set of parameters as in Fig. 15,
along with experimental ARPES data for $\rm Sr_2 Cu O_2 Cl_2 $
from Ref. [2].}
\end{figure}

Taking the same cuprate parameters
as obtained from a recent magnon spectrum fit,\cite{pallab}
and including a small $t''$ term, we have obtained the 
spectral function for key states
and the quasiparticle dispersion along $(\pi,0)$ --- $(0,\pi)$,
as shown in Figs. 15 and 16.
Excellent agreement with the photoemission experiments is obtained with regard to
i) quasiparticle dispersion along $(\pi,0)$ --- $(0,\pi)$,
ii) sharp peaks near $(\pi/2,\pi/2)$ and $(2\pi/3,0)$, and
iii) insignificant (low) spectral weights for $(0,0)$ and $(\pi,0)$ states.
With the same NN hopping term $t=0.35$ eV, our finite-$U$ study yields
$t' \approx -0.09$ eV and $t'' \approx 0.05$ eV,
which are quite smaller than the $t-J$ model values
($t' \approx -0.12$ eV and $t'' \approx 0.09$ eV) of Ref. [3].

\section{Conclusions}
We have obtained one-particle self-energy corrections in the
AF state of the $t-t'-t''$ Hubbard model within a 
physically transparent fluctuation approach involving
the dynamical transverse spin fluctuations.
The full self energy consisting of the inter and intraband contributions,
describes both the quantum correction to ordered moment and ground state
energy due to interband spectral weight transfer,
as well as the intraband, inelastic spin-flip scattering leading to quasiparticle
renormalization due to multimagnon emission and absorption processes
associated with hopping of the mobile carrier. 

In the strong coupling limit,
the intraband self-energy expression is shown to reduce exactly to the SCBA result
for the $t-J$ model, apart from a classical dispersion term 
absent in the $t-J$ model.
That the formally weak-coupling many-body scheme can be carried over
to the strong-coupling limit, indicates its applicability
in the full $U$ range from weak to strong coupling.
By comparing the Hubbard and $t-J$ model results,
we have shown that the classical dispersion term has a significant effect
on quasiparticle energy, mass, and weight, especially with increasing $J$,
and indeed, some features of photoemission experiments on cuprates,
such as vanishing spectral weight for ${\bf k}=(0,0)$,
find natural explanation within the Hubbard model results.

The Hubbard model study provides for a smooth interpolation between
the strong and weak coupling limits. 
The importance of multi-magnon processes
and string excitations in the antiferromagnet,
manifested in pronounced oscillations in spectral function and density of states,
gradually diminishes with decreasing $U$,
and the spectral function becomes increasingly coherent,
the real part of self energy shows a dominant single-magnon contribution,
and the DOS smoothly approaches the free-particle limit.
For the Hubbard model, the one-particle density of states
shows a finite band gap for all $U$ values,
and the ratio $E_g ^* /E_g ^0$
of the renormalized band gap and the classical (mean-field) gap
is found to be nearly 1/2 for a broad range of $U$ values in the
intermediate coupling regime.

The role of fermion scattering states on quasiparticle renormalization
due to fermion-magnon interaction is highlighted by the $t'$-induced asymmetry,
leading to quite different self-energy corrections for the two AF bands.
We find pronounced strong-coupling signature of multi-magnon processes
for the narrow band,
whereas the broad band exhibits significantly weaker renormalization.

For finite $t'$, we find that the AF band gap vanishes
at a critical interaction strength $U_c$.
For $U < U_c$, electron-hole pairs can be spontaneously excited
due to band overlap, which will reduce the sublattice magnetization
and hence the classical gap, and thus further increase the band overlap.
A first-order AF insulator - PM metal transition is therefore expected
at $U=U_c$. 
The $z^2$ dependence of self energy suggests a significantly
stronger suppression of the AF band gap in three dimensions,
yielding a moderate $U_c$ for even smaller $t'$ value.

With the same set of cuprate parameters
as obtained from a recent magnon spectrum fit,\cite{pallab}
excellent agreement with ARPES data for $\rm Sr_2 Cu O_2 Cl_2 $ is obtained,
both with respect to quasiparticle weight as well as dispersion,
effectively providing a unified description of
magnetic and electronic excitations in cuprates.
The $t''$ value required is only half of that obtained within the $t-J$ model
study of Ref. [3], highlighting the role of the finite-$U$
induced classical dispersion term $4J\gamma_{\bf k}^2$.

In a three dimensional system at finite temperature,
thermal excitation of magnons will enhance the
self-energy correction, further decreasing the band gap and leading to,
for $U \gtrsim  U_c$, a temperature driven first-order phase transition.
An exploration of both quantum and thermal phase transitions
from the AF insulator to the PM metal,
as relevant for systems such as $\rm V_2 O_3$, is thus possible
from quasiparticle renormalization due to self-energy correction.
The $t-t'$ Hubbard model also provides a concrete realization of
a stable antiferromagnetic state for finite doping.\cite{karen,stable2}
The doped AF state is characterized by magnon damping due to decay
into particle-hole excitations across the Fermi energy,\cite{affl}
which will have a significant role on quasiparticle damping
through the imaginary part of the fermion self energy.
These consequences of dimensionality,
finite temperature, and doping are presently under study.

\appendix
%\section{Appendix}
\section{Self energy in the AF state}
If $[\Sigma({\bf k},\omega)]$ represents the self-energy matrix
in the two-sublattice basis,
then in terms of the mean-field eigenvalue $E_{\bf k}^0$ 
and eigenvector $|{\bf k}\rangle = \left 
(\begin{array}{c} \alpha_{\bf k} \\ \beta_{\bf k} \end{array} \right )$,
the renormalized Green's function
\begin{eqnarray}
[G({\bf k},\omega)] &=& \frac{1}
{
[G^0({\bf k},\omega)]^{-1} - [\Sigma({\bf k},\omega)]
} \nonumber \\
&=&
\frac{|{\bf k}\rangle \langle {\bf k} | }
{\omega - E_{\bf k}^0 - \Sigma_{\bf k}(\omega)} \; ,
\end{eqnarray}
where expectation value of the self-energy matrix 
\begin{equation}
\Sigma_{\bf k}(\omega) \equiv  \langle {\bf k} | \Sigma({\bf k},\omega) 
| {\bf k} \rangle = (\alpha_{\bf k}^* \;\; \beta_{\bf k}^*) 
\left [ \begin{array}{lr} \Sigma_{AA} & \Sigma_{AB} \\
\Sigma_{BA} & \Sigma_{BB} \end{array}\right ]
\left (
\begin{array}{c}
\alpha_{\bf k} \\ \beta_{\bf k}
\end{array}
\right ) 
\end{equation}
yields the self energy $\Sigma_{\bf k}(\omega)$ for state ${\bf k}$.

Quasiparticle dispersion, weight, and density of states can then be
obtained from Eq. (A1) for the renormalized fermion propagator.
The renormalized quasiparticle energy $E_{\bf k} ^*$ 
is obtained by solving
\begin{equation}
\omega-E_{\bf k}^0 = {\rm Re} \; \Sigma_{\bf k}(\omega) \; ,
\end{equation}
and the quasiparticle weight is given by
\begin{equation}
Z_{\bf k}=\left [1-\frac{\partial}{\partial \omega} {\rm Re}\; 
\Sigma_{\bf k} (\omega) \right]^{-1} _{\omega=E_{\bf k} ^*} \; .
\end{equation}
In the absence of an intersection,
the quasiparticle energy is obtained
from the location of the peak in the spectral function
\begin{equation}
A_{\bf k}(\omega) = \frac{1}{\pi} {\rm Im} \; {\rm Tr} [G({\bf k},\omega)]
= \frac{1}{\pi} {\rm Im} \; \frac{1}{\omega - E_{\bf k} ^0 - \Sigma_{\bf k}(\omega)}
\; ,
\end{equation}
which also yields the density of states 
\begin{equation}
N(\omega) = \frac{1}{\pi} \sum_{\bf k}
{\rm Im} \; {\rm Tr} [G({\bf k},\omega)]
= \sum_{\bf k} A_{\bf k}(\omega) \; .
\end{equation}

\section{Interband contribution}
An additional interband contribution to the self energy is obtained when the
intermediate fermion state ${\bf k-q}$ in Eq. (8) lies in the opposite band.
Absent within the $t-J$ model approach,
this contribution involves interband transfer of spectral weight,
resulting in sublattice magnetization reduction
due to quantum spin fluctuations.
At the one-loop level, this interband contribution
is exactly cancelled to leading order by the
Hartree self energy correction due to spin density
(ordered moment) renormalization.
We also obtain the one-loop correction to ground state energy
as a simple application.

%17
\begin{figure}
%\vspace*{10mm}
\hspace*{-7mm}
\epsfig{figure=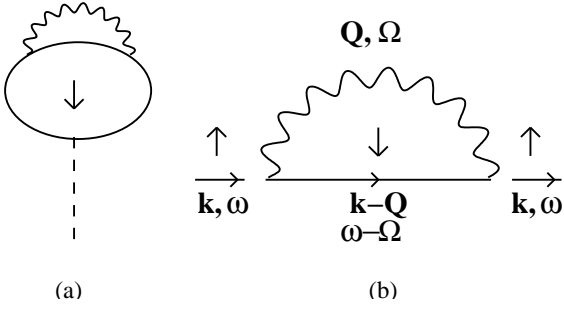,width=75mm,height=40mm}
\vspace{0mm}
\caption{First-order self-energy corrections, representing
(a) correction to the mean-field (Hartree) energy and
(b) mixing between states of the two AF bands.}
\end{figure}

With upper band fermion state $|{\bf k-q} \downarrow \rangle $,
and a retarded magnon pole from Eq. (6),
the interband contribution to the self-energy matrix of Eq. (8) is
\begin{equation}
[\Sigma ' ({\bf k},\omega)] = U^2 \sum_{\bf q} 
\frac{
|\sigma '({\bf k},{\bf q})\rangle
\langle \sigma '({\bf k},{\bf q}) | }
{\omega-\Omega_{\bf q}-E^\oplus_{\bf k-q}
- \Sigma_{\bf k-q} ^\oplus (\omega-\Omega_{\bf q}) + i\eta}
\end{equation}
where, in analogy with Eq. (10), the composite fermion-magnon amplitude
$|\sigma '({\bf k},{\bf q})\rangle = 
\left (\begin{array}{c} v_{\bf q}\; \alpha_{\bf k-q} \\ 
u_{\bf q}\; \beta_{\bf k-q} \end{array} \right )$,
and
\begin{equation}
\Sigma_{\bf k-q} ^\oplus (\omega-\Omega_{\bf q})
= \langle {\bf k-q} |
[ \Sigma ({\bf k-q},\omega-\Omega_{\bf q}) ]
|{\bf k-q} \rangle
\end{equation}
is the expectation value
of the {\em total} self-energy matrix in the upper band state
$|{\bf k-q}\rangle =
\left (
\begin{array}{c}
\beta_{\bf k-q} \\ \alpha_{\bf k-q}
\end{array} \right )$.
Therefore,
for the lower band (hole) state $|{\bf k}\rangle$ in consideration,
the additional interband self energy 
\begin{eqnarray}
\Sigma_{\bf k} ' (\omega) &\equiv & 
\langle {\bf k} | [\Sigma ' ({\bf k},\omega)] | {\bf k} \rangle 
\nonumber \\
&=&
U^2 \sum_{\bf q} 
\frac{ 
(\alpha_{\bf k} v_{\bf q}\; \alpha_{\bf k-q} + 
\beta_{\bf k} u_{\bf q}\; \beta_{\bf k-q} )^2 }
{\omega- \Omega_{\bf q} -E^\oplus_{\bf k-q}
- \Sigma_{\bf k-q} ^\oplus (\omega-\Omega_{\bf q}) } \; .
\nonumber \\
\end{eqnarray}
The above interband contribution transfers spectral weight
from lower to upper band, and together with Eq. (11),
constitutes the full hole self energy,
which describes both the hole renormalization due to
multi-magnon intraband scattering described earlier,
as well as the interband transfer of spectral weight responsible
for density and magnetization quantum corrections.

\subsection{One-loop correction --- Strong coupling limit}
The one-loop self-energy diagrams shown in Fig. 17
represent (a) static correction to the mean-field potential
due to density corrections, 
and (b) mixing between states of the two AF bands.
Green's function correction due to diagram (b)
yields one-loop quantum corrections to spin densities and sublattice
magnetization.\cite{as,quantum}
Spin-$\downarrow$ density increases (decreases) on A (B)
sublattice sites due to transfer of spectral weight from (to) the upper band.
For diagram (a),
these density corrections
of magnitude $\frac{1}{2} \sum_{\bf q} 
\left ( \frac{1}{\sqrt{1-\gamma_{\bf q}^2} } -1 \right )$
yield a Hartree self-energy 
\begin{equation}
[\Sigma_\uparrow ^{(a)}] = 
U [\delta n_\downarrow] = 
U \sum_{\bf q} \frac{1}{2}
\left ( \frac{1}{\sqrt{1-\gamma_{\bf q}^2} } -1 \right )
\left [ \begin{array}{lr} 1 & 0 \\ 0 & -1 
\end{array} \right ]  \; .
\end{equation} 

And for diagram (b) representing mixing,
for a state $|{\bf k} \uparrow \rangle $ in the lower (upper) band,
and the intermediate state $|{\bf k-q} \downarrow \rangle $
in the upper (lower) band,
substituting the strong coupling fermion and magnon ampitudes in Eq. (B1),
we obtain the self energy for the lower band
\begin{eqnarray}
& & [\Sigma_\uparrow ^{(b)} ({\bf k},\omega)]^\ominus
= U^2 \frac{1}{2} \sum_{\bf q}
\frac{1}{\omega-\Omega_{\bf q}-E^\oplus_{\bf k-q} + i\eta}
\nonumber \\
&\times &
\left [ \begin{array}{lr}  
\left (1-\frac{\epsilon_{\bf k-q} ^2}{U^2}\right )
\left (\frac{1}{\sqrt{1-\gamma_{\bf q}^2} } - 1 \right ) & 
\left (\frac{\epsilon_{\bf k-q}}{U}\right )
\left (\frac{-\gamma_{\bf q}}{\sqrt{1-\gamma_{\bf q}^2} } \right ) \\
\left (\frac{\epsilon_{\bf k-q}}{U}\right )
\left (\frac{-\gamma_{\bf q}}{\sqrt{1-\gamma_{\bf q}^2} }  \right ) &
\left (\frac{\epsilon_{\bf k-q} ^2}{U^2}\right )
\left (\frac{1}{\sqrt{1-\gamma_{\bf q}^2} } + 1 \right ) 
\end{array} \right ]  \nonumber \\
\end{eqnarray}
and for the upper band
\begin{eqnarray}
& & [\Sigma_\uparrow ^{(b)} ({\bf k},\omega)]^\oplus =
U^2 \frac{1}{2} \sum_{\bf q}
\frac{1}{\omega+\Omega_{\bf q}-E^\ominus_{\bf k-q} - i\eta}
\nonumber \\
&\times &
\left [ \begin{array}{lr}  
\left (\frac{\epsilon_{\bf k-q} ^2}{U^2}\right )
\left (\frac{1}{\sqrt{1-\gamma_{\bf q}^2} } + 1 \right ) & 
\left (\frac{-\epsilon_{\bf k-q}}{U}\right )
\left (\frac{-\gamma_{\bf q}}{\sqrt{1-\gamma_{\bf q}^2} } \right ) \\
\left (\frac{-\epsilon_{\bf k-q}}{U}\right )
\left (\frac{-\gamma_{\bf q}}{\sqrt{1-\gamma_{\bf q}^2} }  \right ) &
\left (1-\frac{\epsilon_{\bf k-q} ^2}{U^2}\right )
\left (\frac{1}{\sqrt{1-\gamma_{\bf q}^2} } - 1 \right ) 
\end{array} \right ]  \nonumber \\
\end{eqnarray}
These self-energy corrections have the formal structure
of a second-order process involving mixing between states,
and result in energy lowering (raising) for lower (upper) band states.

An exact cancellation to leading order ($U$)
is seen between the Hartree shift
$[\Sigma_\uparrow ^{(a)}]$ of Eq. (B5)
and the mixing contributions
$[\Sigma_\uparrow ^{(b)}]$ of Eqs. (B6) and (B7),
when the self energies $\Sigma_{\bf k} ^\ominus (\omega)$
and $\Sigma_{\bf k} ^\oplus (\omega)$
are evaluated from the corresponding matrices for the lower and upper bands,
with $\omega=E^\ominus _{\bf k}$ and $E^\oplus _{\bf k}$, respectively,
and the surviving terms of order $J$ yield (for the lower band)
\begin{eqnarray}
\Sigma_{\bf k} ^\ominus &=&
\sum_{\bf q}
\left ( \frac{1}{\sqrt{1-\gamma_{\bf q}^2} } -1 \right )
\frac{\epsilon_{\bf k-q} ^2}{U}
-
\frac{\gamma_{\bf q}}{\sqrt{1-\gamma_{\bf q}^2} }
\frac{\epsilon_{\bf k}\epsilon_{\bf k-q}}{U} \nonumber \\
&+& J(1- \sqrt{1-\gamma_{\bf q}^2}) \;.
\end{eqnarray} 

The degeneracy in the classical (mean-field) dispersion along the
magnetic zone boundary $(\epsilon_{\bf k} = 0$) 
is lifted by the first term in Eq. (B7).
The $(\pi/2,\pi/2)$ state is pushed up more
in comparison to the $(\pi,0)$ state,
because $\epsilon_{\bf k-q}^2$ goes like
$(\sin q_x + \sin q_y)^2$ and $(\cos q_x - \cos q_y)^2$, respectively,
and therefore the small $q$ contribution dominates for 
${\bf k}=(\pi/2,\pi/2)$. 

\subsection{Ground-state energy}
Due to mixing of states between the two bands,
$[\Sigma_\uparrow ^{(b)} ({\bf k},\omega)]$ for a lower-band state
yields spectral function in the upper band and vice-versa,
resulting in transfer of spectral weight between the two AF bands.
These modifications to the lower-band spectral function 
result in a quantum correction to the ground state energy,
which we evaluate to first order from the corresponding
Green's function correction.

While the coherent spectral weight in the lower AF band is reduced
by $\frac{1}{2} \sum_{\bf q} 
\left ( \frac{1}{\sqrt{1-\gamma_{\bf q}^2} } -1 \right )$,
an incoherent spectral function with identical integrated weight
is transferred from the upper to the lower band.
Focussing on the latter,
for a state $|{\bf k}\oplus \rangle $,
the first-order correction yields, to leading order, 
\begin{eqnarray}
& & [\delta G^{(1)} _\uparrow ({\bf k},\omega)]_{BB} =
    [\delta G^{(1)} _\downarrow ({\bf k},\omega)]_{AA} \nonumber \\
& = & 
\frac{1}{2} \sum_{\bf q} 
\left ( \frac{1}{\sqrt{1-\gamma_{\bf q}^2} } -1 \right )
\frac{1}{\omega+\Omega_{\bf q} - E^\ominus_{\bf k-q} - i\eta} \; ,
\nonumber \\
\end{eqnarray}
and the corresponding incoherent spectral function is
\begin{equation}
A_{\bf k}^{\rm incoh}(\omega)=
\frac{1}{2} \sum_{\bf q} 
\left ( \frac{1}{\sqrt{1-\gamma_{\bf q}^2} } -1 \right )
\delta(\omega+ \Omega_{\bf q} - E^\ominus_{\bf k-q}) \; .
\end{equation}

The first-order quantum correction $\delta E_{\bf k}^{(1)}$
to lower-band energy can thus be obtained
by simply subtracting the mean-field contribution corresponding to the 
(coherent) spectral weight lost from the lower band,
and adding the energy contribution of the (incoherent) spectral weight
transferred to the lower band.
With $E_{\bf k}^\ominus \approx -\Delta - \epsilon_{\bf k}^2/U$,
we obtain
\begin{eqnarray} 
\delta E_{\bf k}^{(1)} &=&
-E^\ominus_{\bf k}
\frac{1}{2} \sum_{\bf q} 
\left ( \frac{1}{\sqrt{1-\gamma_{\bf q}^2} } -1 \right )
+ \int d\omega \; \omega A_{\bf k}^{\rm incoh} (\omega) \nonumber \\
& = &
- \frac{1}{2} \sum_{\bf q} 
\left ( \frac{1}{\sqrt{1-\gamma_{\bf q}^2} } -1 \right )
\left (\frac{\epsilon_{\bf k-q}^2}{U} - \frac{\epsilon_{\bf k}^2}{U}
+ \Omega_{\bf q} \right )  \; .
\nonumber \\
\end{eqnarray}
Summing over lower-band ${\bf k}$ states for both spins
yields
\begin{equation}
\delta E_g^{(1)} =  \sum_{\bf k} \delta E_{\bf k}^{(1)} 
= -J \sum_{\bf q} \left (1-\sqrt{1-\gamma_{\bf q}^2} \right ) 
\end{equation}
for the quantum correction to ground-state energy,
which is exactly the spin-wave-theory
result for the Heisenberg model.\cite{swa}
The net lowering in ground-state energy is thus a consequence of
the incoherent spectral function in the lower band,
spread over the magnon-energy scale.

\end{document}